\documentclass[]{an}
\usepackage{graphicx}
\usepackage{times}

\begin{document}

\Pagespan{1}{}
 \DOI{This.is/not.aDOI}%

\title{Self-similar Evolutionary Solutions for Accreting Magneto-fluid around a Compact
Object with Finite Electrical Conductivity}

\author{F. Habibi\inst{1}\fnmsep\thanks{Corresponding author:
  \email{f\_habibi@birjand.ac.ir}}
\and  R. Pazhouhesh\inst{1} \and M. Shaghaghian\inst{2}}
\titlerunning{Self-similar Evolutionary Solutions for Accreting Magneto-fluid around a Compact
Object}
\authorrunning{F. Habibi et al.}
\institute{ Department of Physics, Faculty of Sciences, University
of Birjand , Birjand, Iran \and Department of Physics, Science and
Research Branch, Islamic Azad University, Fars, Iran}


\keywords{accretion, accretion disk, magnetohydrodynamic.}

\abstract {%
In this paper, we investigate the time evolution an
 accreting magneto-fluid with finite conductivity. For the case of
 a thin disk, the fluid equations along with Maxwell
equations are derived in a simplified, one-dimensional model that
neglects the latitudinal dependence of the flow. The finite
electrical conductivity is taken into account for the plasma through
Ohm law; however, the shear viscous stress is neglected, as well as
the self-gravity of the disk. In order to solve the integrated
equations that govern the dynamical behaviour of the magneto-fluid,
we have used a self-similar solution. We introduce two dimensionless
variables, $S_0$ and $\epsilon_\rho$, which show the magnitude of
electrical conductivity and the density behaviour with time,
respectively. The effect of each of these on the structure of the
disk is studied. While the pressure is obtained simply by solving an
ordinary differential equation, the density, the magnetic field, the
radial velocity and the rotational velocity are presented
analytically. The solutions show that the $S_0$ and $\epsilon_\rho$
parameters affect the radial thickness of the disk. Also, the radial
velocity and gas pressure are  more sensitive to electrical
conductivity in the inner regions of disk. Moreover, the
$\epsilon_\rho$ parameter has a more significant effect on physical
quantities in small radii.}

\maketitle

\section{Introduction}
Accretion of gas onto compact objects can provide a powerful energy
source for the production of high-energy radiation. When the gas
infalls on a compact object, roughly 10\% of the accreted rest mass
energy could be converted into radiation. Thus, accretion is a
process that can be considerably more efficient than many other
common astrophysical mechanisms, such as nuclear fusion. Because the
removal of the angular momentum process operates on slower
timescales compared to the free-fall time, infalling gas with
sufficiently high angular momentum can form a disk-like structure
around a central compact object, which can be thin or thick
depending on the geometrical shape. Theoretically, thin disks are
well understood, based on a pioneering work by Shakura and Sunyaev
(1973, hereafter SS73). However, for thick accretion disks, no fully
developed model exists, and there remain many theoretical
uncertainties about their structure and stability (Banerjee et al.
1995; Ghanbari \& Abbassi 2004; Ghanbari et al. 2007).

 It is generally believed that magnetic fields have a fundamentally
important role in the physics of accretion disks. Magnetized
accretion disks have been investigated by a number of authors in
several  different contexts: e. g., in relation to X-ray pulsars in
closed binary systems (Ghosh \& Lamb 1978; Aly 1980; Kaburaki 1986),
to bipolar flows and jets from young stellar objects (Pudritz \&
Norman 1983; Kaburaki \& Itoh 1987), and to quasars and active
galactic nuclei (Blandford 1976; Blandford \& Znajek 1977; Macdonald
\& Thorn 1982). A mechanism for angular momentum transport is
another key ingredient in the theory of accretion processes and many
theoretical uncertainties still remain about its nature. A magnetic
field can also contribute to the angular momentum transport. The
angular momentum  is transported by a global magnetic field which
includes both poloidal and toroidal components of the ordered field
(Kaburaki 2000). Additionally, a robust mechanism of the excitation
of magnetohydrodynamical (MHD) turbulence was shown to operate in
accretion disks as a result of the magnetorotational instability
(Balbus \& Hawley 1998; Begelman \& Pringle 2007).

Because plasmas are electrically conducting gases, the influence of
a magnetic field on the gas flow is quite complicated (Frank et
al.1992). In studying magnetized accretion disks from a theoretical
point of view, there are two assumptions about the electrical
conductivity of fluid. The first assumption is that the plasma disk
has infinite conductivity, so that the magnetic fields of compact
stars are completely confined within a certain distance and cannot
penetrate the disk (Davidson \& Osteriker 1973; Lamb et al. 1973;
Aly 1980; Hayakawa 1985; Ogilvie 1997). The other assumption is that
the electrical conductivity must be finite, which is taken into
account through Ohm's law. The importance of finite conductivity was
first pointed out by Ghosh and Lamb (1979), in the interaction of
the accretion disks with the stellar magnetosphere. They showed that
magnetic lines of force can penetrate the accretion disk owing to
the presence of a finite resistivity. An analytical equilibrium
solution, including the conductivity of the plasma for the case of a
thin non-rotating magnetized star accreting matter from a disk, was
obtained by Kaburaki (1986, 1987). He proposed a model for
non-viscous, magnetized accretion disks, and showed that some
features of this model are parallel to those of the standard model
SS73 predicted for viscous, non-magnetic disks. Furthermore, he
pointed out that the inclusion of  finite conductivity is
particularly essential for a disk in the absence of shear viscous,
in order to liberate gravitational energy. Tripathy, Prasanna and
Das(1990, hereafter TPD90) developed this analysis for thick disks
and investigated their stability in the presence of a dipolar
magnetic field due to a non-rotating central object. At the
relativistic limit, Shaghaghian (2011) studied the dynamic of a
stationary disk with finite conductivity around a rotating compact
object with a dipole magnetic field. These studies have shown that
the presence of a magnetic field and its associated finite
conductivity can change the picture of accretion flows.

In this paper, we want to explore how the dynamic of a magneto-fluid
depends on electrical conductivity. Thus, we pursue the approach
adopted by TPD90 for an accreting magneto-fluid surrounding a
non-rotating compact object. The essential difference in our
calculations lies in the inclusion of the time-dependence of
physical variables and the consideration of a thin geometrical
configuration. So, we solve MHD equations for accreting gas that is
self-similar over time. This paper is organized as follows. In
section 2, we first define the general problem and derive the basic
equations for magneto-fluid . Then, we use the self-similar method
to solve the integrated equations that govern the dynamical
behaviour of the accreting gas in section 3. Finally, we present a
summary of the model in section 4.

\section{General formulation}

As mentioned above, we follow the general equations for a
magneto-fluid in a manner similar to TPD90. In his approach, the
Newtonian limit of the relativistic MHD equations are derived for an
axisymmetric magneto-fluid  in a spherical polar coordinate system,
$(r, \theta, \varphi)$, with the origin fixed on the non-rotating
compact object. The accreting gas is highly ionized with finite
electrical conductivity, and the poloidal model is adopted for the
electromagnetic field, in which it has a poloidal component in the
disk ($B_\varphi = E_\varphi = 0$). We also include time dependence
for the matter distribution and electromagnetic fields. Thus, the
Maxwell equations, which are used for MHD approximations, are
obtained in the following form:
\begin{equation}
 J_{r}=  J_{\theta} = 0,
\end{equation}
\begin{equation}
 J_{\varphi}= -\frac{c}{4 \pi r} \bigg[\frac{\partial}{\partial r} \bigg(r
B_{\theta}\bigg)-\frac{\partial B_{r}}{\partial \theta}\bigg]
,\label{basic8}
\end{equation}
\begin{equation}
 J_{t}= -\frac{1}{4 \pi r^2}\bigg[ \frac{\partial}{\partial r}
\bigg(r^2 B_{\theta}V_{\varphi} \bigg)- \frac{r}{sin
\theta}\frac{\partial}{\partial \theta} \bigg(sin\theta B_{r}
V_{\varphi} \bigg)\bigg],\label{basic9}
\end{equation}
\begin{equation}
\frac{\partial}{\partial \theta} \bigg(r sin\theta B_{\theta}
\bigg)+\frac{\partial}{\partial r} \bigg(r^2 sin\theta B_{r}
\bigg)=0,\label{basic10}
\end{equation}
\begin{equation}
\frac{\partial E_{r} }{\partial \theta} +\frac{\partial}{\partial r}
\bigg(r E_{\theta} \bigg)=0,\label{basic20}
\end{equation}
\begin{equation}
\frac{\partial}{\partial t} \bigg(r sin\theta B_{\theta}
\bigg)=0,\label{basic19}
\end{equation}
\begin{equation}
\frac{\partial}{\partial t} \bigg(r^2 sin\theta B_{r}
\bigg)=0.\label{basic11}
\end{equation}
Moreover, the electrical conductivity is taken into account through
Ohm's law
\[
  J^i = \sigma F^i_ k  u^k,
\]
where  $J^i$,  $F_{ik}$ and $u^k$ are the current density, the
electromagnetic field tensor and the four velocity vectors,
respectively.  The electrical conductivity of fluid, $\sigma$, is
assumed to be constant throughout the disk. By neglecting toroidal
fields, Ohm's law yields
\begin{equation}
E_{r}=\frac{B_{\theta} V_{\varphi}}{c},\label{basic12}
\end{equation}
\begin{equation}
E_{\theta}=-\frac{B_{r} V_{\varphi}}{c},\label{basic13}
\end{equation}
\begin{equation}
J_{\varphi}=-\frac{\sigma}{c} \bigg( B_{\theta} V_{r}-B_{r}
V_{\theta} \bigg),\label{basic14}
\end{equation}
\begin{equation}
J_{t}=-\frac{\sigma}{c} \bigg(E_{\theta} V_{\theta}+E_{r} V_{r}
\bigg)=\frac{J_{\varphi} V_{\varphi}}{c}.\label{basic15}
\end{equation}
The equations (\ref{basic8}) and (\ref{basic14}) are two different
definitions for the azimuthal component of the current density.
Their consistency  yields
\begin{equation}
\frac{\partial}{\partial r} \bigg(r B_{\theta} \bigg) -
\frac{\partial B_{r}}{\partial \theta} = \frac{4 \pi\sigma r}{c^2}
\bigg( B_{\theta} V_{r}-B_{r} V_{\theta} \bigg).\label{basic18}
\end{equation}
An admissible solution set for magnetic field that satisfies the
equations (\ref{basic10}), (\ref{basic19}) and (\ref{basic11}) is
\begin{equation}
B_r = - B_1 r^{k-1}sin^{k-1}\theta cos \theta,\label{basic30}
\end{equation}
\begin{equation}
B_\theta =  B_1 r^{k-1}sin^{k}\theta ,\label{basic31}
\end{equation}
where $k$ and $B_1$ are the constant values. From the above
equations, we can see that the poloidal magnetic field does not
depend explicitly on time. It has a similar configuration to the
stationary disk (TPD90), which represents constant field lines
parallel to the meridional plane. Substituting equations
(\ref{basic30}) and (\ref{basic31}), the consistency relation
(\ref{basic18}) reduces to
\begin{equation}
V_r+V_\theta cot \theta = \frac{(k-1)c^2}{4\pi \sigma r sin^2
\theta}. \label{trans32}
\end{equation}
In the following, we consider a  geometrically thin configuration,
which means that the vertical thickness of the disk is sufficiently
small, that we can assume $\theta = \pi/2$ and neglect terms
associated with the $\theta$-velocity ($V_\theta = 0$) and any
$\theta$ dependence. The viscosity of the plasma has been ignored,
since our purpose is analysing the role of electrical conductivity
in an accreting gas. Also, the magnetic and electric stresses do not
play a role in angular momentum transfer, due to the absence of the
toroidal components of the fields. The angular momentum transfer,
therefore, occurs through the finite electrical resistivity of the
plasma, $\eta = \frac{c^2}{4\pi \sigma}$, (Kaburaki 1987). For the
sake of simplicity, the self-gravity of the disk is ignored in
comparison with the gravitation of the central compact object.
Consequently, the motion of plasma is governed by the continuity
equation
\begin{equation}
\frac{\partial \rho}{\partial
t}+\frac{1}{r^2}\frac{\partial}{\partial r}(r^2 \rho
V_r)=0,\label{basic1}
\end{equation}
and the momentum equations
\begin{equation}
\frac{\partial V_r}{\partial t}+V_r \frac{\partial V_r}{\partial
r}+\frac{MG}{r^2}-\frac{V_\varphi^2}{r}+\frac{1}{\rho}\frac{\partial
P}{\partial r}+\frac{B_\theta}{\rho}
\frac{J_\varphi}{c}=0,\label{basic2}
\end{equation}
\begin{equation}
\frac{\partial V_\varphi}{\partial t}+V_r \frac{\partial
V_\varphi}{\partial r}+\frac{V_\varphi V_r}{r} =0,\label{basic3}
\end{equation}
where $P$, $\rho$, $V_r,V_\varphi$ denote the gas pressure, density,
radial and azimuthal components of plasma velocity, respectively.
Applying the thin disk approximation, the radial velocity is
attained by the equation (\ref{trans32})
\begin{equation}
V_r = \frac{(k-1)c^2}{4 \pi \sigma r}.\label{Vr}
\end{equation}
As a result, the radial velocity has the same form in steady (TPD90)
and unsteady solutions. This is because the time derivative does not
appear in the consistency relation. The equations of motion,
(\ref{basic1}), (\ref{basic2}) and (\ref{basic3}), are a set of
non-linear partial differential equations, which cannot be solved
analytically. Therefore, we search for solutions that describe the
temporal change of physical quantities in such a way that the change
of each quantity at any moment of time is similar to that of the
others.

\section{Self-similar solutions}
\subsection{Analysis}
The equations of motion can be transformed into a set of ordinary
differential equations by a temporal self-similar approach. The
technique of self-similar analysis is familiar from its wide range
of applications in the full set of equations of MHD in many research
fields of astrophysics. A similarity solution, although constituting
only a limited part of problem, is often useful in understanding the
basic behaviour of the system.  In a self-similar formulation, we
introduce a similarity variable $\xi$ as
\begin{equation}
\xi = \frac{r}{r_0(t)},\label{self1}
\end{equation}
wherein $r_0(t)$ follows a time-dependent power-law relation, as
$r_0(t)=a t^{n}$ and $a,n$ are constants that are used to make $\xi$
dimensionless. So, physical quantities (functions of $r$ and $t$)
are assumed as unknown time-dependent coefficients and functions of
the dimensionless variable, $\xi$. We can determine  the unknown
time-dependent coefficients in a way that satisfies the basic
equations. The resulting algebraic equations, which are only a
function of the dimensionless variable $\xi$, can be solved
analytically or semi- analytically. Therefore, we consider physical
quantities in the following forms:
\begin{equation}
V_r(r,t)=v_0(t) V_r(\xi),\label{self2}
\end{equation}
\begin{equation}
\rho(r,t)=\rho_0(t) \rho(\xi),\label{self3}
\end{equation}
\begin{equation}
V_\varphi(r,t)=w_0(t) V_\varphi(\xi),\label{self4}
\end{equation}
\begin{equation}
 P(r,t)=p_0(t) P(\xi),\label{self5}
\end{equation}
\begin{equation}
B_\theta(r,t)=b_0(t) B(\xi).\label{self6}
\end{equation}
Also, we assume the electrical conductivity as
\begin{equation}
\sigma = S_0 \sigma_0(t).\label{self7}
\end{equation}
Here, we define $S_0$ as a conductivity dimensionless coefficient
that is used to study the effect of electrical conductivity and as a
free parameter.  The equation (\ref{self7}) is necessary to create
equations in a dimensionless form, since electrical conductivity has
the dimension $time^{-1}$. Now, the equations (\ref{basic31}) and
(\ref{Vr}) can be rewritten as
\begin{equation}
B_\theta(r,t) = \bigg(B_1 r_0(t)^{k-1}\bigg)\xi^{k-1},
\end{equation}
\begin{equation}
V_r(r,t) = \bigg(\frac{c^2}{4\pi
\sigma_0(t)r_0(t)}\bigg)\frac{k-1}{S_0 \xi},
\end{equation}
\textsf{wherein}
\begin{equation}
b_0(t) = B_1 r_0(t)^{k-1},\qquad B(\xi) = \xi^{k-1},\label{B1}
\end{equation}
\begin{equation}
v_0(t) = \frac{c^2}{4\pi \sigma_0(t)r_0(t)},\qquad V_r(\xi) =
\frac{k-1}{S_0 \xi}.\label{vrx}
\end{equation}
By substituting equations (\ref{self1})-(\ref{self7}) into equations
(\ref{basic1})-(\ref{basic3}), the following relations are obtained
\begin{equation}
r_0(t) = (GM)^{1/3} t^{2/3},\label{self8}
\end{equation}
\begin{equation}
v_0(t) = (GM)^{1/3} t^{-1/3},\label{self9}
\end{equation}
\begin{equation}
p_0(t)/\rho_0(t) = (GM)^{2/3} t^{-2/3},\label{self10}
\end{equation}
\begin{equation}
w_0(t) = (GM)^{1/3} t^{-1/3},\label{self11}
\end{equation}
\begin{equation}
b_0(t)^2/\rho_0(t) = 4 \pi (GM)^{2/3} t^{-2/3},\label{self12}
\end{equation}
\begin{equation}
\sigma_0(t) = \frac{c^2}{4\pi}(GM)^{-2/3}t^{-1/3}.\label{self13}
\end{equation}
It should be mentioned that the current component $J^{\varphi}$ in
equation (\ref{basic2}) has been inserted by equation
(\ref{basic14}).  The above results imply that each physical
quantity retains a similar spatial shape as the flow evolves, but
the radius of the flow increases in proportion  to $t^{2/3}$.
\textsf{ In self-similar space,} the time-dependent behaviour of the
velocities is proportional to $t^{-1/3}$, namely, they decrease with
time. Also, these relations show  $p_0(t)$ and $b_0(t)$ are
dependent on the behaviour of $\rho_0(t)$. Our results are in
agreement with the previous findings for such time-dependent systems
(Ogilvie 1999; Khesali \& Faghei 2008, 2009). Moreover, self-similar
solutions indicate that electrical conductivity is scaled with time
as $t^{-1/3}$. This is a logical consequence of the dimensional
consideration perspective.

For specifying the time dependence of $\rho_0(t)$, and then $p_0(t)$
and $b_0(t)$, we also assume a time-dependent power-law relation for
density as
\begin{equation}
\rho_0(t) = R_0 t^{\epsilon_\rho},\label{self14}
\end{equation}
where $R_0$ is an arbitrary coefficient with the dimension of
density $\times$ time $^{-\epsilon_\rho}$. In order to evaluate the
exponent $\epsilon_\rho$, we employ the mass accretion rate $\dot
M$:
\begin{equation}
\dot M = -4 \pi r^2 \rho V_r.\label{self20}
\end{equation}
Just like the equations (\ref{self2})-(\ref{self7}) for the mass
accretion rate, we can write:
\begin{equation}
\dot M(r,t) = \dot M_0(t) \dot M(\xi),\label{self15}
\end{equation}
under transformations of equations (\ref{self1}), (\ref{self2}) and
(\ref{self3}), equation (\ref{self15}) becomes
\begin{equation}
\dot M(r,t) = [ r_0(t)^2\rho_0(t)v_0(t)] [
-4\pi\xi^2\rho(\xi)V_r(\xi)],\label{self16}
\end{equation}
which implies
\begin{equation}
\dot M_0(t) = r_0(t)^2\rho_0(t)v_0(t),\label{self17}
\end{equation}
\begin{equation}
\dot M(\xi)= -4\pi\xi^2\rho(\xi)V_r(\xi).\label{self18}
\end{equation}
 By substituting equations
(\ref{self8}), (\ref{self9}) and (\ref{self14}) into the equation
(\ref{self17}), we obtain:
\begin{equation}
\dot M_0(t) = R_0 (GM) t^{\epsilon_\rho+1}.\label{self19}
\end{equation}
The above relation shows that when $\epsilon_\rho = -1$, $\dot
M_0(t)$ is constant and decreases in $\epsilon_\rho < -1$.
Accordingly, we can obtain a set of solutions for $\epsilon_\rho\leq
-1 $. Now, it is possible to derive the  constant values of $B_1$
and $k$ in equation (\ref{B1}), by using the equations
(\ref{self12}) and (\ref{self14}). Thus, we obtain
\begin{equation}
B_1 = (4 \pi R_0)^{\frac{1}{2}} (GM)
^{\frac{1}{3}(2-k)},\label{self19}
\end{equation}
and
\begin{equation}
 k=\frac{1}{2}(1+\frac{3}{2}\epsilon_\rho).\label{k}
\end{equation}
Therefore, self-similar solutions show that the value of $k$ is
restricted by the $\epsilon_\rho$ parameter. This means that the
spacial behavior of the magnetic field, in current time dependent
solution, is determined by the $\epsilon_\rho$ parameter. However,
as pointed out by TPD90, the magnetic field lines inside the disk do
not depend on the value of $k$. Moreover, the magnetic moment of the
disk, $B_1$, is proportional to the matter distribution in the disk,
while in the steady state, this parameter was obtained by boundary
conditions and proportional to the surface magnetic field of the
compact object. Also, by employing the equation (\ref{k}), we find
$V_r = - c^2/16\pi \sigma r$ for the maximum value of $k=-1/4$,
whereas $V_r = - c^2/4\pi \sigma r$ for the maximum value of $k=0$
in the steady solution. This indicates that although the radial
dependence of infall velocity is same as the stationary disk, but
its magnitude is changed when we consider the effects of
continuously growing central mass.

\subsection{Transformed basic equations}

Substituting  self-similar solutions (\ref{self1})-(\ref{self7})
into the equations of motion, a set of ordinary differential
equations is obtained as
\begin{equation}
\frac{d \rho(\xi)}{d\xi}\bigg[V_r(\xi)-
\frac{2}{3}\xi\bigg]+\rho(\xi)\bigg[\frac{d
V_r(\xi)}{d\xi}+2\frac{V_r(\xi)}{\xi}+\epsilon_\rho\bigg]=0,\label{trans1}
\end{equation}
\[
-\frac{1}{3}V_r(\xi)+\bigg[V_r(\xi)-\frac{2}{3}\xi\bigg]\frac{d
V_r(\xi)}{d\xi} - \frac{V_\varphi ^2(\xi)}{\xi}+ \frac{1}{\xi^2}
\]
\begin{equation}
+ \frac{1}{\rho(\xi)} \frac{d P(\xi)}{d\xi} + S_0
\frac{B(\xi)^2}{\rho(\xi)} V_r(\xi)= 0,\label{trans2}
\end{equation}
\begin{equation}
\bigg[V_r(\xi) -\frac{2}{3}\xi\bigg] \frac{d V_\varphi(\xi)}{d\xi}+
\bigg[\frac{
V_r(\xi)}{\xi}-\frac{1}{3}\bigg]V_\varphi(\xi)=0.\label{trans3}
\end{equation}
 The equations (\ref{trans1}) and (\ref{trans3}) can be solved analytically by introducing $V_r(\xi)$
 into them. Thus, the azimuthal velocity and the density
distribution are
\begin{equation}
\label{result4} V_\varphi(\xi) =
\frac{l}{\xi}\bigg[\xi^2+\frac{3}{2}\frac{(1-k)}{ S_0}\bigg]^{1/4},
\end{equation}
\begin{equation}
\label{result3}
 \rho(\xi) = \frac{1}{\xi}\bigg[\xi^2+\frac{3}{2}\frac{(1-k)}{S_0}\bigg]^{\frac{1}{2}(1+\frac{3}{2}\epsilon_\rho)}.
\end{equation}
Here,  $l$ is defined as the angular momentum parameter. It should
be mentioned that the radial dependence of azimuthal velocity and
density is different from steady solutions. Furthermore, in current
solution, the azimuthal velocity is controlled by the electrical
conductivity, as well as the radial velocity.

\begin{figure*}
\includegraphics[width=150mm]{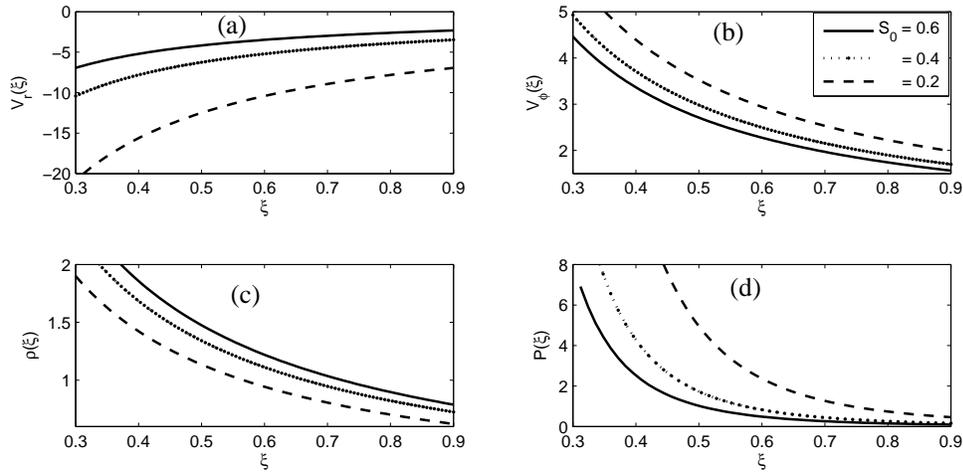}
\caption {The typical behavior of radial and azimuthal components of
velocity, density and pressure  as a function of dimensionless
variable $\xi$ for $\epsilon_\rho=-1, n = 1$ and some values of
$S_0$ .}\label{figs}
\end{figure*}

Finally, with the aid of the similarity functions $B(\xi),
V_r(\xi)$, $V_\varphi(\xi)$ and $\rho(\xi)$, a simple differential
equation is obtained for the gas pressure:
\[
 \frac{dP(\xi)}{d\xi} =
\frac{1}{\xi}\bigg[\xi^2
+\frac{3}{2}\frac{(1-k)}{S_0}\bigg]^{\frac{1}{2}(1+\frac{3}{2}\epsilon_\rho)}\bigg[\frac{1}{3}\frac{(1-k)}{S_0\xi}-\frac{1}{\xi^2}
\]
\begin{equation}
+\frac{(1-k)^2}{S_0^2\xi^3}+\frac{l^2}{\xi^3}\bigg(\xi^2+\frac{3}{2}\frac{1-k}{S_0}\bigg)^{1/2}
\bigg] -(1-k)\xi^{2k-3}.
\end{equation}
This equation can be easily integrated by numerical methods. As a
result, our solutions are sensitive to the $S_0$ and $\epsilon_\rho$
parameters.  The $S_0$ parameter indicates the role of the
electrical conductivity in the dynamics of the disk, and the
$\epsilon_\rho$ parameter shows density changes with time.  In
fig.~\ref{figs}, we present our results for $\epsilon_\rho = -1$,
the mass accretion rate of which is independent of time, and some
values of $S_0$. As this figure clearly shows, the radial inflow and
azimuthal velocities become faster when the flow is inwards. The
flow has differential rotations. An increase in $S_0$ slows down
velocities and leads to an increase in density and a decrease in gas
pressure. The gas pressure is a descending function of $x$, as well
as of density distribution. Moreover, fig. \ref{figs}c shows
electrical conductivity's effect on the radial thickness of the
disk; specifically, with increasing $S_0$ , the radial thickness of
the disk increases. Also, as can be seen, the radial velocity and
gas pressure are more sensitive to electrical conductivity in the
inner regions of the disk. The behaviour of the radial velocity and
density with respect to the electrical conductivity is qualitatively
consistent with the results of TPD90.

Here, it is useful to compare the our results for a non-viscous disk
with viscous disks. In viscous disks,  the angular momentum is
transported by the turbulent viscosity and the time-dependence of
disk flow is controlled by the size of the viscosity, $\alpha$
(Frank et al 2002). In our model for a non-viscous disk, as
mentioned before, the transfer of angular momentum, which guarantees
the successive accreting of matter, is carried by the electrical
resistivity. The electrical resistivity has the same units as
kinematic viscosity and we define the size of electrical resistivity
as $\eta_0 = 1/S_0$ which can be considered corresponding to the
viscous dimensionless parameter, $\alpha$. We can also claim that
the time-dependence of disk flow is controlled by the size of the
electrical resistivity $\eta_0$. Moreover, previous studies on the
viscous disks (Akizuki \& Fukue 2006; Khesali \& Faghehi 2008)
indicated that by increasing $\alpha$ parameter, the radial infall
velocity increases, on the contrary, the density distribution
decreases. Our self-similar solutions also show the $\eta_0$
parameter has similar effect on the radial velocity and density.
\begin{figure*}
\includegraphics[width=150mm]{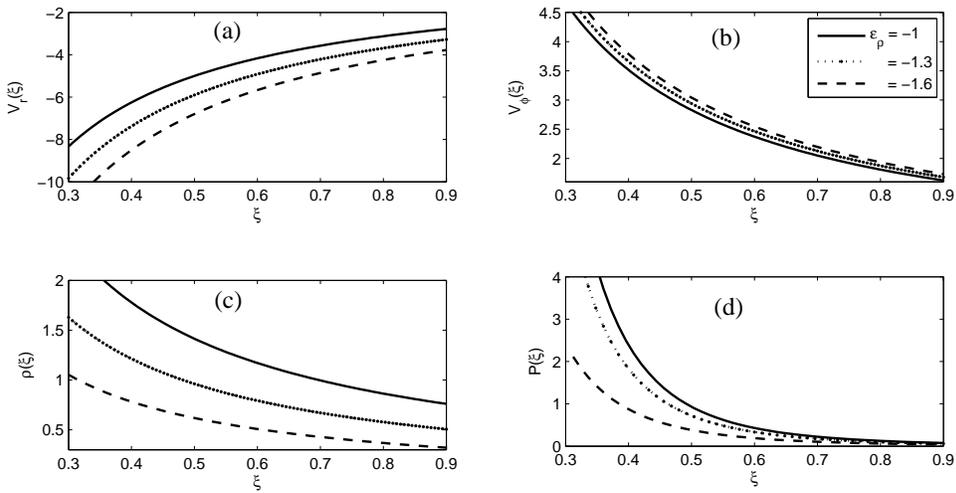}
\caption {Same as for fig.1 for $S_0=0.6$ and different values of
$\epsilon_\rho$}\label{fige}
\end{figure*}
\begin{figure*}
\includegraphics[width=150mm]{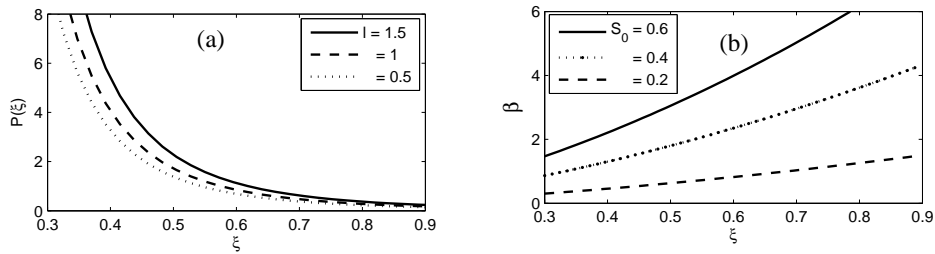}
\caption { Panel (a): the typical behavior of gas pressure for
different values of $n$, constant values are  $S_0=0.6$ and
$\epsilon_\rho=-1$. Panel (b): the $\beta$ parameter for
$\epsilon_\rho=-1$ and different values of $S_0$.}\label{figkb}
\end{figure*}
\begin{figure*}
\includegraphics[width=150mm]{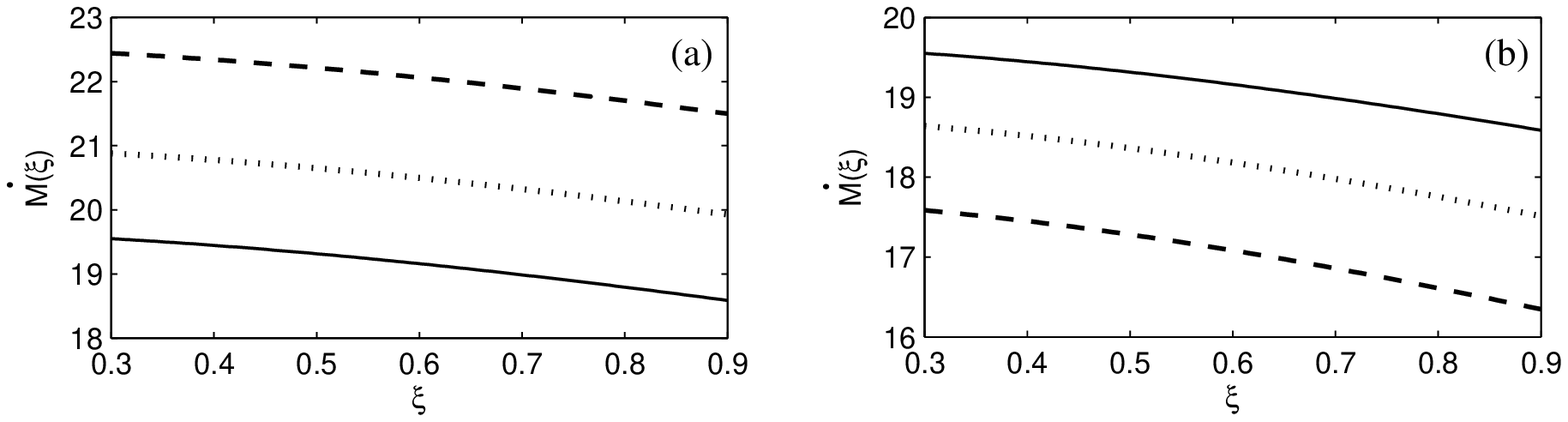}
\caption {Mass accretion rate as a function of dimensionless
variable $\xi$. Panel (a): for $\epsilon_\rho=-1$, the solid lines
represent $S_0 = 0.6$, the dotted lines represent $S_0 = 0.55$, and
the dashed lines represent $S_0 = 0.5$. Panel (b): for $S_0=0.6$,
the solid lines represent  $\epsilon_\rho = -1$, the dotted lines
represent $\epsilon_\rho = -1.1$, and the dashed lines represent
$\epsilon_\rho = -1.2$.}\label{figm}
\end{figure*}

We plot fig.\ref{fige} similarly to fig.\ref{figs} for $S_0=0.6$ and
some values of $\epsilon_\rho$. This figure shows that  both the
radial and the azimuthal components of the velocity increases with
increases in the absolute value of $\epsilon_\rho$. This is contrary
to the behaviour of the density and gas pressure, which reduce as
the absolute of $\epsilon_\rho$ becomes larger. Moreover, the
solutions show that the $\epsilon_\rho$ parameter is important at
small radii. Panel (c) of fig. \ref{fige} indicates that the radial
thickness of the disk decreases with increase of the absolute value
of $\epsilon_\rho$.

The effect of angular momentum parameter $l$ on the distribution of
pressure is demonstrated in fig.\ref{figkb}(a). It shows that for
higher values of $l$ , the pressure increases. This result is
qualitatively consistent with the results of TPD90. On the other
hand, for an accreting magneto-fluid, the effective pressure has
included gas pressure and magnetic pressure
\begin{equation}
\bar P = P + P_{mag} = P + \frac{B^2}{8\pi}.
\end{equation}
We can show the importance of magnetic field pressure in comparison
with the gas pressure by $\beta$ parameter, which is defined as the
ratio of the magnetic pressure to the gas pressure. By using the
self-similar solutions (\ref{self1})-(\ref{self7}), the $\beta$
parameter takes the form
\begin{equation}
\beta(r,t) = \frac{B_{\theta}^2(r,t)/8\pi}{ P(r,t)} = \frac{1}{2}
\frac{B(\xi)^2}{ P(\xi)}.
\end{equation}
The above relation shows that the $\beta$ parameter is independent
of time. In other words, the fact that the time-dependent behaviour
of the magnetic and gas pressures is similar limits the
self-similarity solution. In contrast, physical quantities with the
same physical dimension have similar behaviour (Khesali \& Faghei
2009). In fig.\ref{figkb}(b), we plot the $\beta$ parameter as a
function of $\xi$ for several values of $S_0$. This figure indicates
that the dominant pressure in the outer region of the disk is the
magnetic pressure; this result is consistent with observed young
stellar object disks (Aitken et al. 1993; Greave et al. 1997). The
$\beta$ parameter varies by $S_0$; that is, it increases by
increasing the $S_0$ parameter, but this variation is important at
larger radii.

By substituting equations (\ref{vrx}) and (\ref{result3}) into
equation (\ref{self18}), the function of $\dot M(\xi)$ is obtained
as
\begin{equation}
\dot M(\xi)= \frac{4\pi(1-k)}{S_0}
\bigg[\xi^2+\frac{3}{2}\frac{(1-k)}{S_0}\bigg]^k.
\end{equation}
We plot $\dot M(\xi)$ as a function of $\xi$ for some values of
$S_0$ and $\epsilon_\rho$ in fig.~\ref{figm}.  $\dot M(\xi)$ is very
sensitive to the values of $S_0$ and $\epsilon_\rho$. Therefore,
these parameters are selected in smaller ranges. The figure displays
the accretion rate is a function of position, and increases with
decreasing radii, while in the steady state, it is a constant. As
can be seen in fig.~\ref{figm}(a) , when $S_0$ becomes smaller, the
mass accretion rate increases. Such behaviour is similar to the
behaviour of radial velocity. This is because radial velocity has
more sensitivity to electrical conductivity than density (see
equation \ref{self20}). By contrast, the mass accretion rate
decreases as the absolute of $\epsilon_\rho$ becomes larger.

\section{Conclusions}

The main aim of this manuscript was to examine the effect of
electrical conductivity on the dynamic of accreting magneto-fluids.
To this end, a self-similar solution has been derived for the
equations of time-dependent accretion in a one-dimensional model.
 The plasma has been considered to be resistive with
azimuthal and radial velocities. We have restricted ourselves to
flow accretions in which self-gravitation and shear viscosity are
negligible. In self-similar space, the electrical conductivity
coefficient ,$\sigma$, which is assumed to be constant throughout
the disk, is scaled with time as $t^{-1/3}$ and a dimensionless free
parameter, $S_0$, is remained that demonstrates the size of
electrical conductivity. In absent shear viscosity, both angular
momentum transfer and energy dissipation in the flow is undertaken
by the electrical resistivity whose size is specified by the $\eta_0
= 1/S_0$ parameter. The $\eta_0$ parameter can be considered
corresponding to $\alpha$ parameter in standard model SS73.

The basic equations of dimensionless have been solved analytically.
We obtained solutions parameterized by the temporal changes of
density, $\epsilon_\rho$, and the conductivity dimensionless
coefficient, $S_0$. The behaviour of physical quantities is
determined for $\epsilon_\rho = -1$, where the mass accretion rate
is independent of time, and also $\epsilon_\rho<-1$ which decreases
with time. The solutions indicate that the $S_0$ parameter has a
significant effect on the gas pressure and radial infall velocity in
the inner regions of the disk. The radial thickness of the disk
increases with greater $S_0$ parameter. Moreover, the solutions show
the $\epsilon_\rho$ parameter is important in the inner regions of
the disk. By increasing the absolute value of $\epsilon_\rho$, the
radial thickness of the disk decreases and the disk become
compressed.

The previous studies of time dependent systems with infinite
conductivity reveal the radial behaviour of  physical quantities was
different from the results achieved by those who considered
stationary states (Khesali \& Faghei 2008, 2009). Our result,
however, indicate that the radial dependence of the inflow velocity
is similar to the solution of the stationary state which is reported
by TPD90. This is a consequence of the presence of finite
conductivity, which was taken in to account through Ohm law. We have
also shown that the ratio of the magnetic pressure to the gas
pressure is a function of position, and increases when moving
outward. This means that the magnetic field is more important at
large radii. This ratio becomes larger by increasing the electrical
conductivity of flow. Moreover, the solution shows that the mass
accretion rate varies by radii and time and is very sensitive to
values of $S_0$ and $\epsilon_\rho$.

In our model, the latitudinal dependence of the physical quantities
is ignored, but our understanding of the disk's properties could be
improved by a discussion of two-dimensional models, in which one
could explicitly treat both the radial and meridional structures of
the accretion flow, and in which the thickness of the disk could be
considered. In future studies, we will develop our model for thick
disks.

\section{Acknowledgment}
We are like to appreciate the referee for his/her thoughtful and
constructive comments to improve this paper.

\end{document}